\documentclass[epj]{svjour}  
\usepackage{graphicx}
\usepackage{epsfig}
\usepackage{subfigure}
\usepackage{amsmath}
\usepackage{rotating}
\usepackage{amsfonts}
\usepackage{amssymb}  
\psfigdriver{dvips}

\begin{document}
\title{Linear response in aging glassy systems, intermittency  
and the Poisson statistics of record fluctuations.} 
\author{Paolo Sibani} 
\institute{Institut for Fysik og Kemi, SDU, DK5230 Odense M, Denmark}  
\date{\today} 
\abstract{ We study    the intermittent behavior  of  the energy decay and linear magnetic response
of a  glassy system during isothermal aging   after a deep thermal quench 
using the  Edward-Anderson spin glass model as a paradigmatic example. 
The large intermittent changes in the two observables are found to
   occur in a correlated fashion  and   through    irreversible  bursts, 
`quakes', which punctuate  reversible and equilibrium-like   fluctuations of zero average. 
The temporal distribution of the quakes it foun  to be
a Poisson distribution with an  average  growing logarithmically on   time,
indicating that the quakes are triggered by record sized fluctuations. 
As  the drift of an aging system  is to a good approximation   subordinated to the quakes, 
simple analytical expressions (Sibani et al. Phys Rev B 74, 224407, 2006)
are available  for the time and age 
dependence of the average response and average energy.
These expressions are shown to capture the time dependencies of  the EA  simulation results.  
Finally, we argue that  whenever the changes  of the linear response function and of its
 conjugate   autocorrelation function  follow from the same intermittent events 
a fluctuation-dissipation-like relation  
can   arise  between the two in off-equilibrium aging. 
} 
  
\maketitle
 
\section{Motivation}  
 In noise spectra from mesoscopic aging systems, 
 reversible  fluctuations  are punctuated  by  rare and   large, so called intermittent, 
events~\cite{Kegel00,Weeks00,Bissig03,Buisson03,Buisson04,Cipelletti05}, which  
arguably  signal     switches   from one metastable  configuration to 
another~\cite{Crisanti04,Sibani05}. The intermittent  events usually 
appear as an exponential tail in the Probability Density Function (PDF) of the 
fluctuations. The  central part
of the PDF  describes equilibrium-like behavior with its  zero-centered Gaussian shape. 
Exploiting this information can     adds    new twists  to 
long debated issues as the    multiscale nature of glassy dynamics
and the associated memory behavior. 

A  \emph{record dynamics}  scenario for aging~\cite{Sibani05,Sibani93a,Sibani03}   
builds on  two  main assumptions:  \emph{(i)}  record sized   energy fluctuations within 
metastable domains trigger \emph{irreversible}  intermittent events, or \emph{quakes};  
\emph{(ii)}   these, in turn, control all
significant  physical changes,   e.g. heat release,  configurational decorrelation  
and   linear magnetic response. In brief, all important changes   are considered to be 
\emph{subordinated} to the quakes, with the latter  triggered by energy fluctuations of \emph{record} size.
Combining the information from the  equilibrium-like fluctuations
  with non-thermal properties, e.g.  irreversible   energy losses, leads to 
testable predictions  for the   time and temperature 
dependencies of the fluctuation spectra which have been confirmed in  a 
number of cases~\cite{Sibani05,Anderson04,Sibani04a,Sibani06,Oliveira05,Sibani06b,Sibani06a}.
 
\begin{figure*} 
$
\begin{array}{cc}
\includegraphics[width=0.47\linewidth,height= 0.47\linewidth]{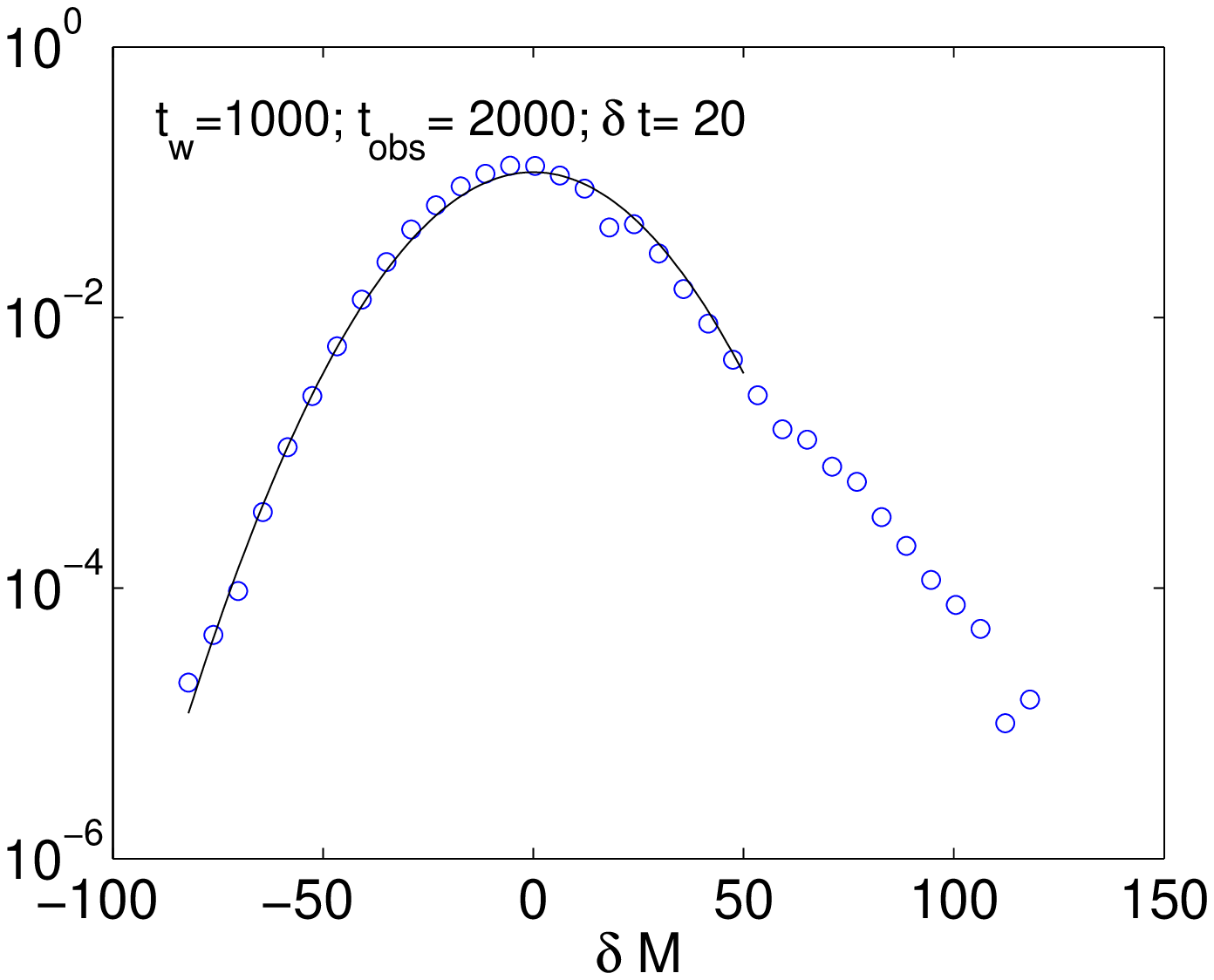}  &  
\includegraphics[width=0.47\linewidth,height= 0.47\linewidth]{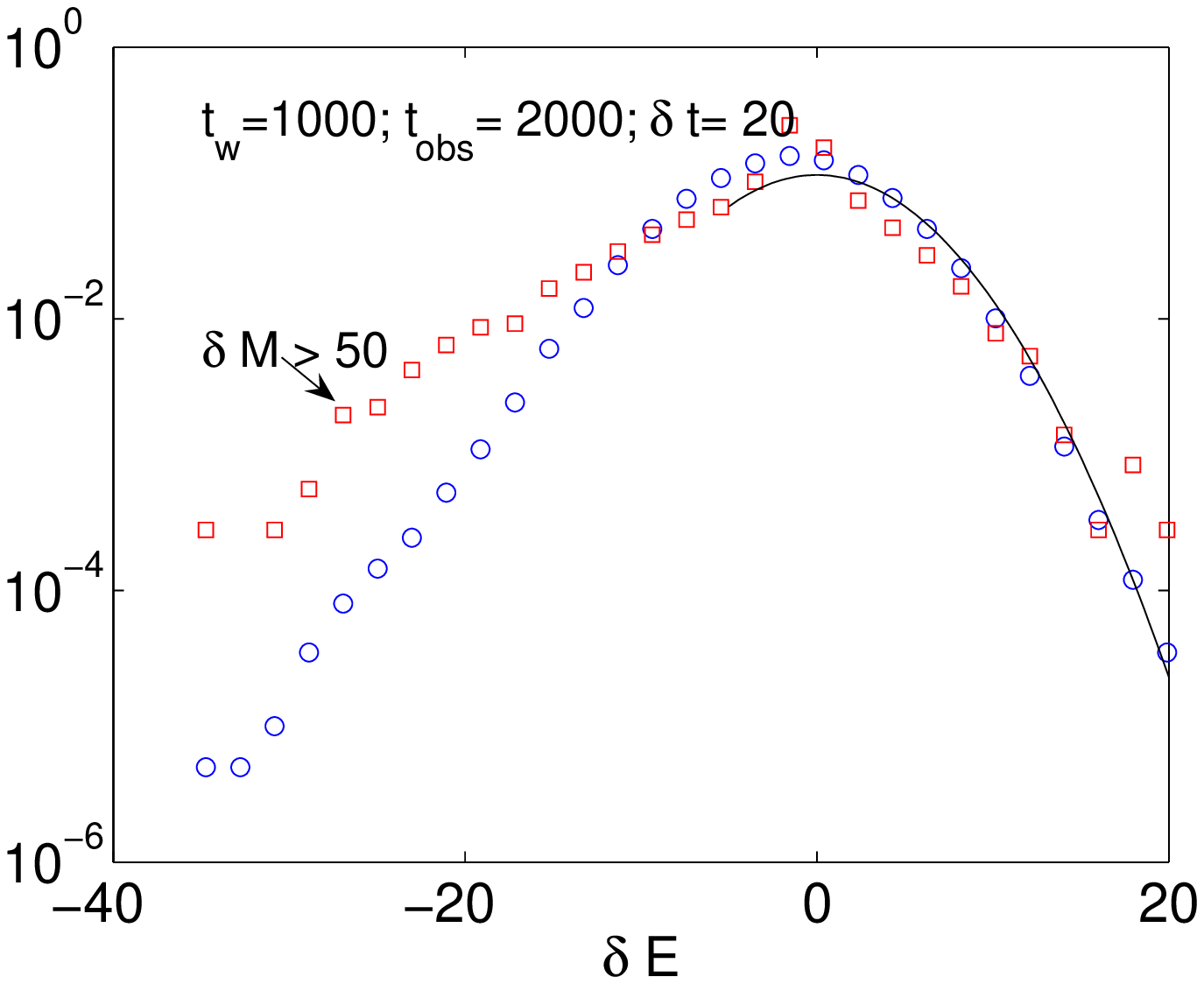}
\end{array} 
$ 
\vspace{0.5cm} 
\caption{(Color on line) The left panel shows  the PDF (blue circles) of the magnetization changes
$\delta M$ occurring over  intervals of length $\delta t = 20$ during  isothermal simulations
at temperature $T=0.4$.
The  statistics  is based on  $2 \cdot 10^5$ data points obtained in the age  interval $[1000,3000]$
using  $2000$ independent trajectories. 
 The  zero average Gaussian (black line)  fits  data points with  $-80 < \delta M < 50$.     
 The right panel shows the PDF of the energy changes (blue circles),
 and a  Gaussian fit of the positive energy fluctuations (black line). The red squares pertain to
 the  subset of energy changes   occurring in the immediate vicinity of large ($\delta M> 50$ )
 magnetization changes. All data are obtained from the same set of trajectories
 as those shown in the left panel.  
}
\label{intermittency1}
\end{figure*} 
While  macroscopic (average) linear response functions
have long been the tool of choice for probing aging in magnetic 
systems,~\cite{Lundgren83,Alba86,Zotev03,Rodriguez03,Suzuki03}, 
the  predictions of record dynamics  have so far only been tested
for   experimental Thermoremanent Magnetization (TRM) data. 
 spin glass data~\cite{Sibani06a}. 
Simulations  offer certain advantages over  experiments:   it is possible to 
simulate an instantaneous quench, simply by choosing a `random', i.e. high
temperature,  initial condition. Notably, an instantaneous
initial quench leads to the  `full aging'~\cite{Rodriguez03},
 scaling behavior   simply  explained    in    record  dynamics; secondly, the statistical analysis is 
considerably  simplified when  thousands of independent traces can be generated; 
lastly,    temporal  correlations between the
intermittent changes of the magnetization and the energy can be extracted.

The Edwards-Anderson (EA) spin-glass model  used in this work is a 
paradigmatic   example of an aging system.
Its   macroscopic (average)   magnetic response and autocorrelation decay 
 have been thoroughly investigated~\cite{Andersson92,Rieger93,Kisker96,Picco01}
 and its    mesoscopic fluctuations properties have    attracted  some
recent  attention~\cite{Castillo03,Sibani05,Sibani06}.
Here,  data from extensive simulations  are analyzed  with  focus  on 
 the intermittency of the energy and Zero Field Cooled Magnetization (ZFCM)  fluctuations 
We  confirm that   quakes   carry the net drift of 
the energy~\cite{Sibani05}
and  correlate strongly  with the large intermittent
magnetization fluctuations  carrying  the net change of the linear response.  
The  idea~\cite{Sibani03,Sibani05} that quakes have a Poisson
distribution with a \emph{logarithmic}  time dependence is derived from record statistics
and is central to the theory. To verify it empirically,  we consider the temporal statistics 
of the difference  of 'logarithmic waiting times' 
$\log(t_q)-\log(t_{q-1})$, where $t_q$ marks the occurence of the $q$'th quake  
in a given trace.
For the above Poisson  distribution,   these  
logarithmic differences  are exponentially distributed. 
 General mathematical arguments  lead to eigenvalue expansions  for the
 dependence of the  average energy and linear response on the 
 number of quakes, and then, via the subordination hypothesis,  to power-law expansions for the time dependences
 of the same quantities~\cite{Sibani06,Sibani06a}. The 
expansions are  tested below  against the EA model numerical simulation results.
The  origin of approximate off-equilibrium Fluctuation-Dissipation like
relations is discussed in the last section from the point of view of 
 record dynamics. 

Finally, a notational issue:  as  in ref.~\cite{Sibani06a},  the variable  $t$  
generically  denotes  the  time elapsed from the
initial quench, i.e. the  system age.    The external field is switched on at time 
$t=t_w$, and  $t_{obs} = t-t_w$ denotes  the observation
time, a quantity  called $t$  in  refs.~\cite{Sibani05,Sibani06} and in many experimental papers
e.g. ~\cite{Vincent96,Rodriguez03}. Unless otherwise stated, we denote the \emph{average} 
energy and magnetization by $\mu_{e}$ and $\mu_{ZFCM}$, and  reserve the symbols $e$ and  $M$ for the
corresponding  fluctuating quantities  measured in the simulations. The simulation temperature
is denoted by $T$. 
 
\section{Model and Simulation methods} \label{results} 
\begin{figure*} 
$
\begin{array}{cc}
\includegraphics[width=0.47\linewidth,height= 0.47\linewidth]{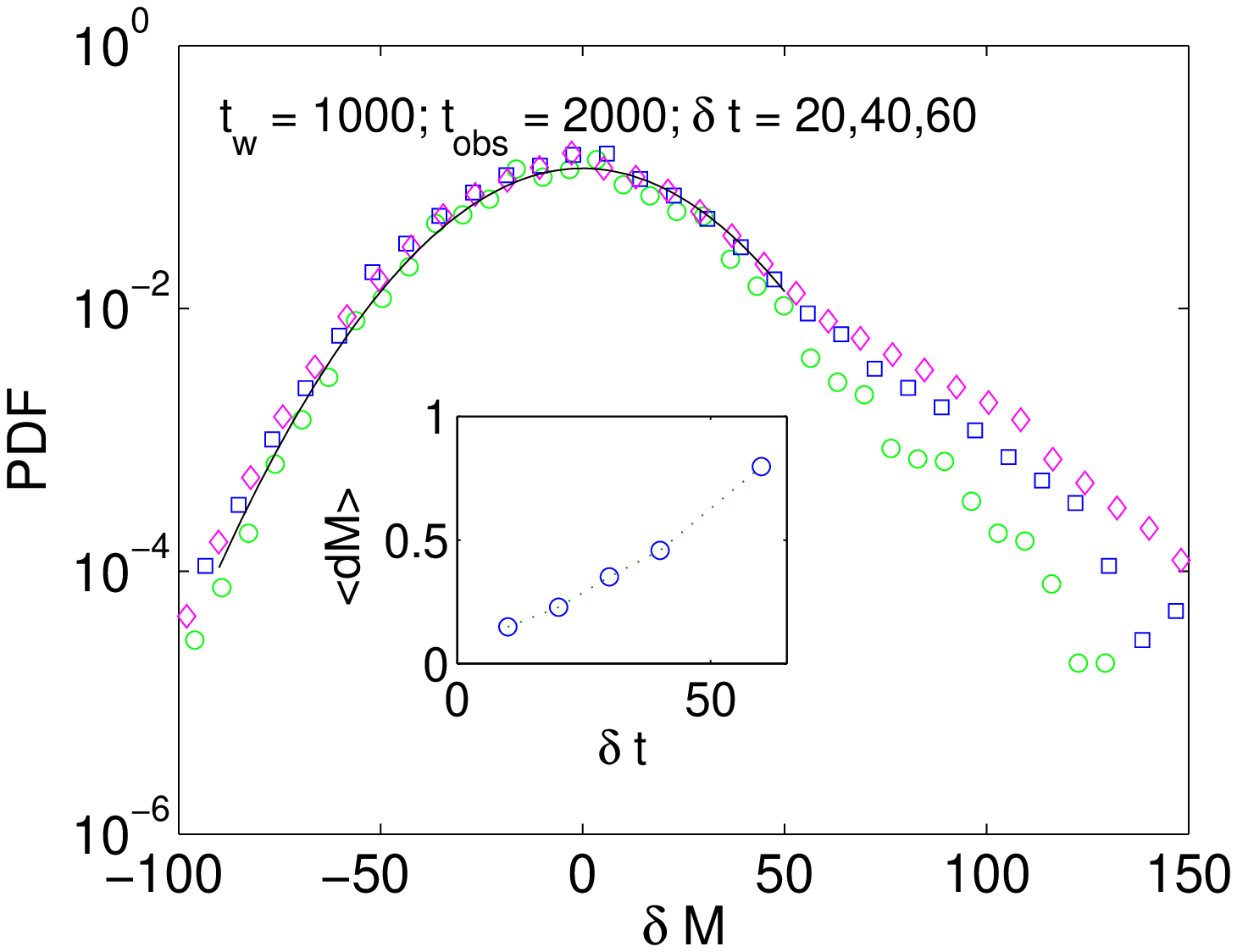}  &  
\includegraphics[width=0.47\linewidth,height= 0.47\linewidth]{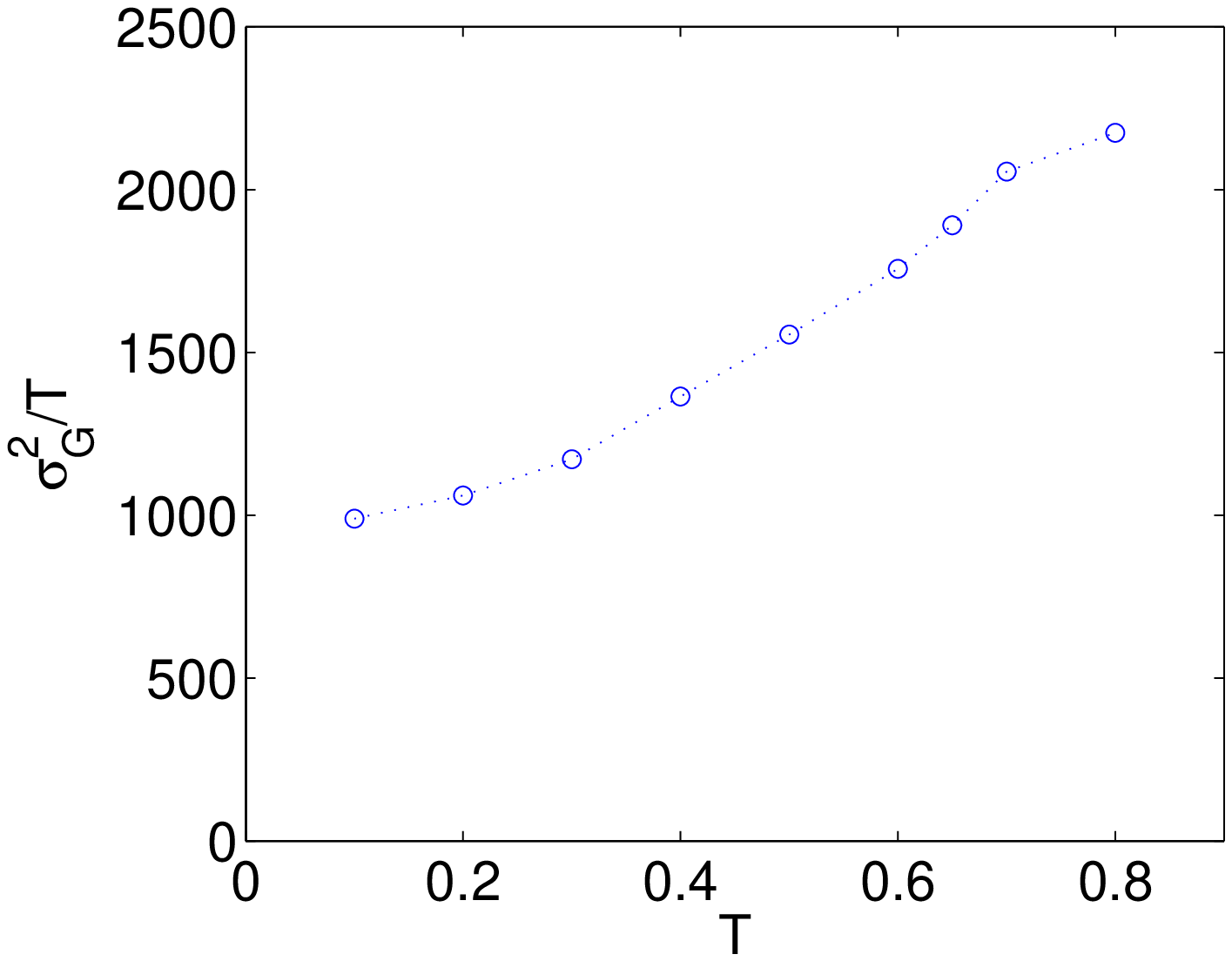}
\end{array} 
$ 
\vspace{0.5cm} 
\caption{(Color on line) The left panel  depicts the PDF of the magnetization changes
$\delta M$ over intervals of length $\delta t$. The simulation  temperature is $T=0.4$.  
Green circles, blue squares and magenta diamonds
  correspond to $\delta t=20,40$ and $60$, respectively. 
  The data are collected in the interval $[1000,3000]$.
  In order of  increasing $\delta t$,
the  statistics is based on 
 $10^5$, $.5 \cdot 10^5$ and $.33 \cdot 10^5$ data points.
The full line is a  fit of  all  data with $-80 < \delta M < 50$ 
to a Gaussian centered at zero. For  $\delta M>50$ the PDF
 clearly  deviates  from a Gaussian shape. The intermittent tail
 grows with  $\delta t$ and fully determines the average change in the ZFC magnetization.   
 The insert shows $dM$, averaged over $t_{obs}$, versus $\delta t$. 
  The right panel shows the variance of the (fitted) Gaussians as a function 
 of the reduced temperature $T/T_g$. 
}
\label{intermittency}
\end{figure*} 
In the  Edwards-Anderson model,   $N$  Ising variables, $\sigma_i = \pm 1$, 
 are placed on  a cubic lattice with  toroidal boundary conditions.
Their  interaction energy    is  
\begin{equation}
E = -\frac{1}{2} \sum_{i,j}^N \sigma_i \sigma_j J_{ij} - H \sum_i^N \sigma_i,
\label{model}
\end{equation}
  \begin{figure}[t!]  
\includegraphics[width=0.95\linewidth,height= 0.85\linewidth]{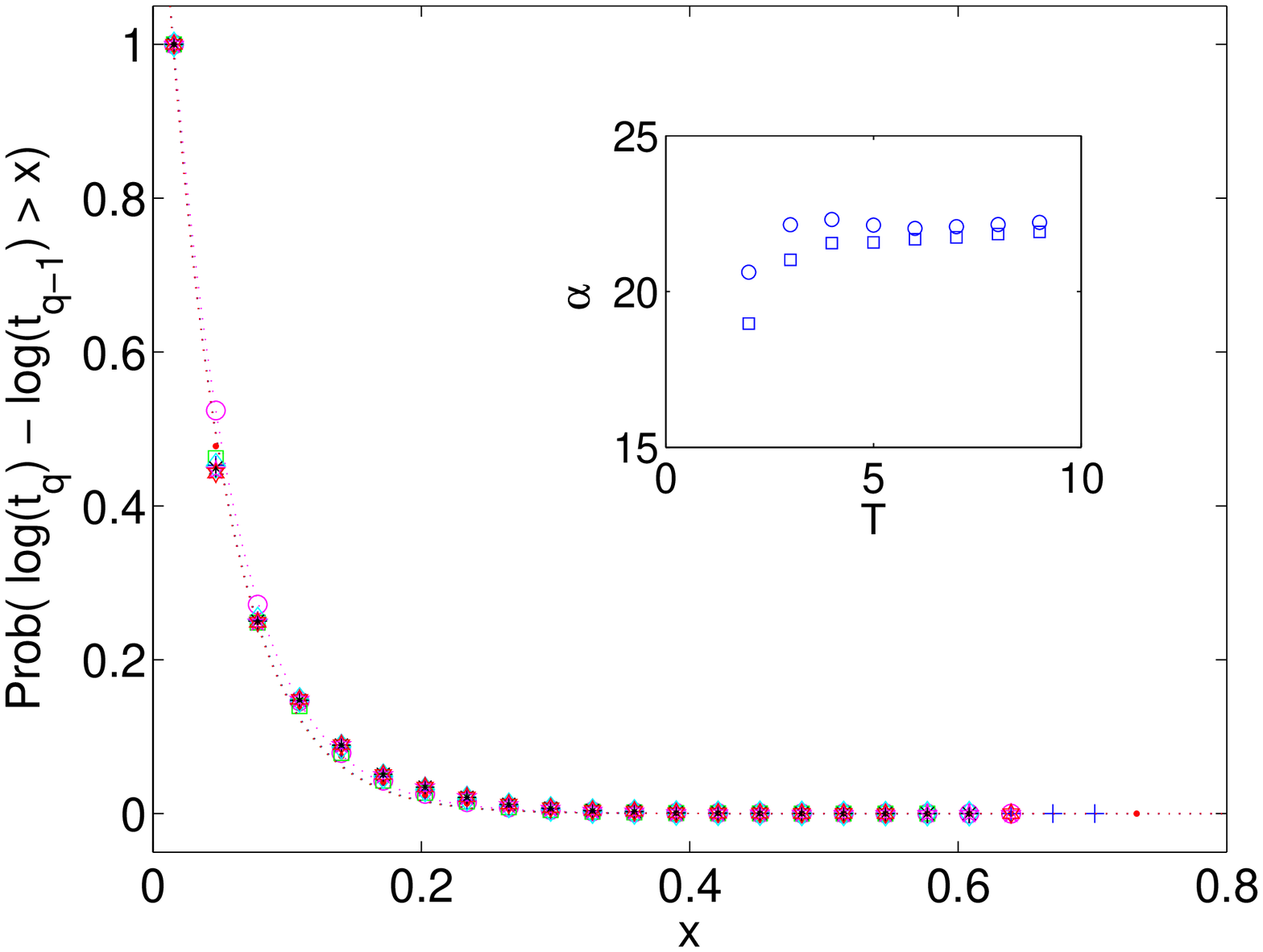}     
 \vspace{0.5cm} 
\caption{(Color on line) 
 For eight  different 
temperatures $T=0.2, 0.3 \ldots 0.9$   
 (circles, dots, squares, plusses, diamonds,stars,  pentagrams, and hexagrams)
the estimated  cumulative probability  Prob$(\log t_q/t_{q-1} > x)$ is plotted 
 versus $x$, with  $t_q$, $q=1,2 \ldots$  denoting  the 
times  at which the    quakes  occur. The quakes are identified as 
described in the main text using the filtering parameter value $f=7$.
The dotted lines are  fits   to  the  exponential  $ \exp(-\alpha x)$.
The  collapse  of the eight data sets demonstrates  that the temperature
dependence   is very weak, except  at the lowest $T$. In the insert,
 the fitted   $\alpha$ values are  plotted versus $T$. The circles
pertain to the data shown in the main plot, and the squares pertain  to 
corresponding data obtained  using  $f=8$ as a filter parameter.
}
\label{logtimeratios}
\end{figure}   
where $\sum_i \sigma_i = M$ is the magnetization 
and $H$ is  the  magnetic field. The
non-zero elements of the  symmetric
interaction matrix $J_{ij}$ 
connect  neighboring sites on the lattice. The interactions   are 
 drawn from a Gaussian distribution with zero average and unit 
variance, a choice  setting  the scale for both  temperature
and  magnetic field.   

In the  simulations,  we use system size   $N=16^3$  
and collect the   statistics of energy and magnetization changes
  from either several thousands of  independent 
 trajectories, each corresponding to a different  realization  of the
 $J_{ij}$'s. 
After the initial quench,   aging procedes  isothermally in zero field
 untill   time $t=t_w$, where a  small magnetic field $H=0.1$ is 
 instantaneously turned on.   Isothermal simulations lasting up to  age $t=100 \, t_w$ 
are  carried out for $t_w = 50, 100, 200, 500 $ and $1000$ at 
 temperatures $T=0.1,0.2, \ldots 0.8$.   
The simulation engine utilizes  the 
 Waiting Time Method~\cite{Dall01,Sibani06b}, a rejectionless
or `event driven' algorithm which  operates   with an 
`intrinsic' time, loosely corresponding   to a sweep  of the Metropolis
algorithm. Unlike the Metropolis algorithm, the WTM
can  follow spatio-temporal patterns on very short time scales,   a     
property advantageous   for  intermittency studies.

\section{Simulation  results} 
In this Section, simulation results are presented together with some relevant 
theoretical considerations.  The  first subsection deals with the  distribution of  
energy and magnetization   changes, $\delta E$ and $\delta M$, occurring  over
 small intervals    $\delta t $. The  Probability Density Function (PDF) of these quantities 
 characterizes  intermittency in 
the EA model in a simple and direct way. In the second subsection,  
the  quakes  are (approximately) indentified within each data stream, and the   
temporal aspects of their distribution are studied.  
 In the third and last subsection, the average linear response and average energy obtained
 from the simulations are compared to analytical   predictions  which are based on 
 the temporal distribution of the quakes. 
\subsection{Energy and magnetization fluctuation statistics}
For a first   statistical description of  the energy and the  ZFCM  
 fluctuations, energy and magnetization changes $\delta E$ and $\delta M$ occurring over  
 small intervals    $\delta t $ are   sampled during the time  interval  $[1000,3000]$. 
The left panel of Fig.~\ref{intermittency1}  shows 
 the  PDF of the magnetic fluctuations
(blue circles) on a log scale. The (black) line  is the  fit  
to a zero-centered Gaussian obtained using data  within the interval $-80< \delta M < 50$.
The distribution of the remaining large positive magnetization changes 
is seen to have  exponential character. The same  runs are used
to collect the   PDF of the energy fluctuations,  shown in
 the   right panel of  Fig.~\ref{intermittency1} (lower  data set, blue circles). 
Again, we see a combination of  a zero centered Gaussian and an intermittent  
  exponential tail. 
The  full line is a fit to the  zero centered Gaussian
in the interval $-5<\delta E< 20$. 

As intermittent fluctuations  are rare,
the   temporal correlation  between  intermittent  energy and ZFC magnetization changes 
is best observed  using  conditional PDF's (upper curve, red squares). 
The conditional PDF   shown only   
  includes those $\delta E$ values   which     either fall 
 in the same  or in the  
preceding small interval of width $\delta t$ as   magnetic fluctuations above the  threshold   $\delta M = 50$.
In the left panel  of the figure,  the threshold  is seen to be near  
the boundary between   reversible and intermittent magnetization changes.  
Note how the Gaussian parts of   the full and of the conditional  PDFs  nearly
coincide, while the corresponding  intermittent  tails differ. The tail of the 
conditional PDF    is strongly enhanced (roughly,
between $10$ and $100$ times) when only  fluctuations  near  a large 
  magnetization change are included in the count. A similar  situation is  observed
at any  low temperatures.  

Figure~\ref{intermittency}   details  the dependence of the PDF 
of the magnetization fluctuations  on  $\delta t$ and $T$: 
The  three PDFs shown in the left panel of  Fig.~\ref{intermittency} as green circles, 
blue squares and magenta diamonds are obtained in the
interval $[1000,3000]$, for $T=0.4$ and 
for  $\delta t=20,40$ and $60$, respectively. 
 The black  line is a zero centered Gaussian  fitted to the PDF  in the
 range  $-80 < \delta M< 10$.  The    fit is obtained  by optimizing 
 the variance of the Gaussian distribution, $\sigma_G^2$. The latter quantity 
 should not  be confused with the variance  of the full distribution, which 
 is much larger and heavily influenced  by  the tail events excluded from
  the Gaussian fit. 
 Importantly, since the Gaussian part of the PDF is independent 
of $\delta t$     quasi-equilibrium   ZFCM fluctuations  
at $T=0.4$ can be treated as  uncorrelated for times larger than $\delta t=20$.
Secondly, since the Gaussian is centered at zero,   any
 changes in the average   magnetization  are exclusively 
 due  to the  intermittent events. Similar conclusions   
were  reached for the  spin-glass   TRM magnetization,~\cite{Sibani06a}
 and  for the energy outflow in the EA model,~\cite{Sibani05}  
 and   in  a p-spin   model with no quenched randomness.~\cite{Sibani06b} 
The  insert  describes the  dependence on $\delta t$ of the 
magnetization change $\langle \delta M \rangle$    averaged 
over the observation  interval $[1000,3000]$. E.g. the  data point 
corresponding to $\delta t = 20$   
depicts  the average of the lowest   PDF  plotted in  the main figure.  
The right panel of fig.~\ref{intermittency}  shows a plot of 
the  ratio   $\sigma^2_G/T$ versus $T$ for the Gaussian part of the 
magnetic fluctuations.  As  the  Fluctuation-Dissipation 
Theorem (FDT) applies to the equilibrium-like part of the dynamics, 
the  ordinate can be interpreted as the (gedanken) linear magnetic susceptibility 
of metastable configurations.   
The  overall shape of the  curve  is reminiscent of the experimental $T$ dependence  
 of  the ZFCM below $T_g$.~\cite{Nordblad87} 
\subsection{Temporal distribution  of intermittent events}
The  PDF of  intermittent fluctuations in the EA model  
  scales  with  the ratio $\delta t/t$, where $t$ is the   age of the system at the 
beginning of the sampling interval~\cite{Sibani05}.   The  rate of quakes  accordingly 
 decays as the inverse of the age, in agreement  with the claim that the 
the number of independent quakes  in an interval $(t,t')$  has a  
 Poisson distribution with  average $\langle n_I \rangle = \alpha  \log(t/t')$  
 The   parameter   $\alpha $ characterizing the average    is
 interpreted as the number of thermalized domains 
contributing  in parallel to the fluctuation statistics. This quantity 
is expected to be    temperature independent 
and linearly  dependent on    system size~\cite{Sibani06b}. 
Using that the differences
 $\tau_q = \log(t_q)-\log(t_{q-1}) = \log(t_q/t_{q-1})$  
are independent random numbers, all exponentially distributed with the same average
$\langle \tau_q \rangle  = 1/\alpha$,  identifying the times 
$t_q$, $q=1,2,\ldots$ at which  the quakes 
 occur allows one to check the  temporal aspects of statistics (as opposed
 to the distributionof quake sizes). 
 
 The identification  entails  some  challanges of 
 statistical and patter-recognition nature.  
 The   residence times $t_q - t_{q-1}$ expectedly have a broad distribution~\cite{Sibani03}, which,  in connection 
with a finite sampling time creates a negative sampling bias on large $t_q$ values~\cite{Anderson04},
even with the quakes   unabiguously   identified. 
Secondly, correlations may well be present 
between large and closely spaced  spikes in the signal which  collectively
 represent a change of attractor. 
 Last but not least,  identification of the $t_q$'s within a trace must rely 
 on the negative sign of the fluctuations   and on their  'sufficiently large'  size.  
I.e. by  definition    sufficient   energy 
must  be released  to  make the reverse process highly  unprobable  within an  observation 
span  stretching up to $2t$ for a quake which occurs at $t$.~\cite{Sibani03} 
 A related  and  generic   property of aging systems 
 (also directly accessible by an intermittency analysis~\cite{Sibani05})
 is that   reversible  equilibrium-like fluctuations dominate the dynamics on time scales shorter than 
the age $t$.  
A simple  empirical criterion for 
distinguishing   reversible fluctuations 
from  irreversible quakes   uses  an upper bound on  the probability,
 $P_{rev}(t,\delta E)$ that  at least one   thermal   energy fluctuation of  positive sign  be 
 among   the     $ t/\delta t$   observations. For small $\exp(-\delta E/T)$, this happens
with probability $P_{rev}(t,\delta E) \approx t/\delta t c \exp(-\delta E/T)$, where the 
positive number $c$   is  unknown. 
  A negative  energy  fluctuation is  labeled as a quake if 
 its   absolute value  $|\delta E|$   satisfies 
  $P_{rev}(t,|\delta E|) < 10^{-n}$,  where $n$ is a  positive number. For a thermal 
  fluctuation which occurs   at age  $t$,  
this criterion produces   the     inequality  
\begin{equation}
\frac{|\delta E|_{\rm quake}}{T} > f + \log t; \quad \mbox{\rm where}  \quad f =  n \ln(10) + \ln (c/\delta t). 
\label{criterion} 
\end{equation}
The formal  definition of the  `filter'  parameter  $f$  on the right hand side of the
inequality contains the unknown parameter 
 $c$ as well as a    free parameter $n$.  We  must therefore   
treat  $f$ as a free parameter   regulating how strict a filter the  fluctuations must pass to 
qualify as  quakes.    
   
For temperatures $T=0.2, 0.3 \ldots 0.9$, a  range spanning most of the EA  spin-glass phase, 
 simulations were performed in the interval $t \in (200,10200)$.
Within each trajectory,  quakes are identified by applying Eq.~\ref{criterion}, and 
the quantities $\tau_q = \log(t_q/t_{q-1})$  are  calculated and  binned. The  procedure 
is  repeated for $2000$ independent trajectories for each value of the temperature.  
The magnetic field values  
  $H=0.1$   for $t > t_w = 200$  and zero otherwise, the same as    in all   other simulations,
  were used for simplicity. Note however that, as also demonstrated by Fig.~\ref{big_e},
   the  magnetic field  has no discernible effects  on the  energy statistics.  
The standard deviation, $\sigma(i)$, of the empirical    probability of a  point falling  in the $i$'th bin 
 is   $\sigma(i) = \sqrt(p_{th}(i)/N_{obs})$, where $p_{th}(i)$ is the  corresponding (unknown)
   theoretical 
probability. $N_{obs}$ is the   number of $\tau_q$ values collected. 
Replacing  $p_{th}$   with the corresponding  empirical probability,
yields an  estimate for $\sigma(i)$. The   empirical cumulative distribution 
Prob$(\tau_q = \log(t_q/t_{q-1})>x)$ is  fitted to the  exponential $b e^{\alpha x}$, with  
fit  parameters obtained  by mimizing 
the  sum of the square differences to the  data points, each   weighted  by the 
reciprocal of the  (estimated) variance of the data point. The prefactor $b$ has no 
physical significance. It  would be 
unity, where not for the fact that the lowest $x$ value, for which the ordinate 
is equal to one,  is located at half  bin size, rather than at zero.    
To check the influence of   $f$ on the form of the distribution,   all the  simulations were
done twice, using  the values $f=7$ and $f=8$. The results  shown in the main 
panel of Fig.~\ref{logtimeratios} 
pertain to the $f=7$ case. The results for $f=8$ are very similar, i.e. 
both cases produce  an   exponential distribution. The scale parameter  $\alpha$  
lacks  any  significant  $T$ dependence,   except for the
lowest temperature, and seems only weakly  $f$  dependent.
  From the statistics, the number of domains 
 can be estimated as  $\alpha \approx 20$. Dividing the total number of spins   
 by $\alpha$   an order of magnitude estimate is obtained 
  for  the  number of spins in a thermalized domain.   
The  corresponding  linear size comes out  around  $6$ spins, a   
  figures which compares well   to  the range of domain sizes observed in  domain growth   
 studies of  the EA model~\cite{Rieger93}.
 The scale invariance of the energy landscape implict in the $T$ independence of 
 $\alpha$ is to a large degree confirmed. It is expected that the largest deviations be found
 at the lowest temperatures, where the discreteness of the energy spectrum begins to make itself felt.  

\subsection{Average energy and linear response} 
\begin{figure*}
$
\begin{array}{cc}
\includegraphics[width=0.45\linewidth,height= 0.4\linewidth]{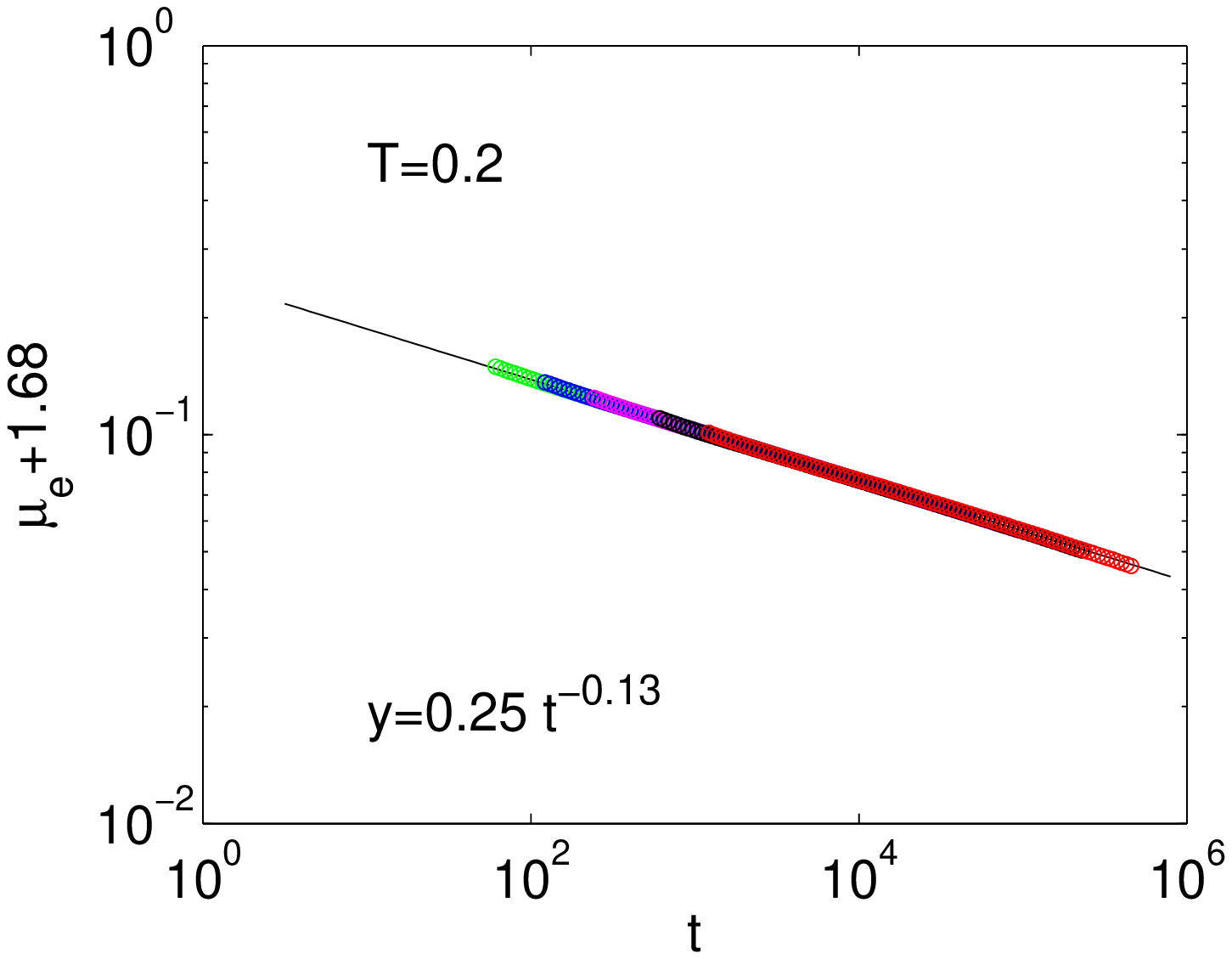}  &  
\includegraphics[width=0.45\linewidth,height= 0.4\linewidth]{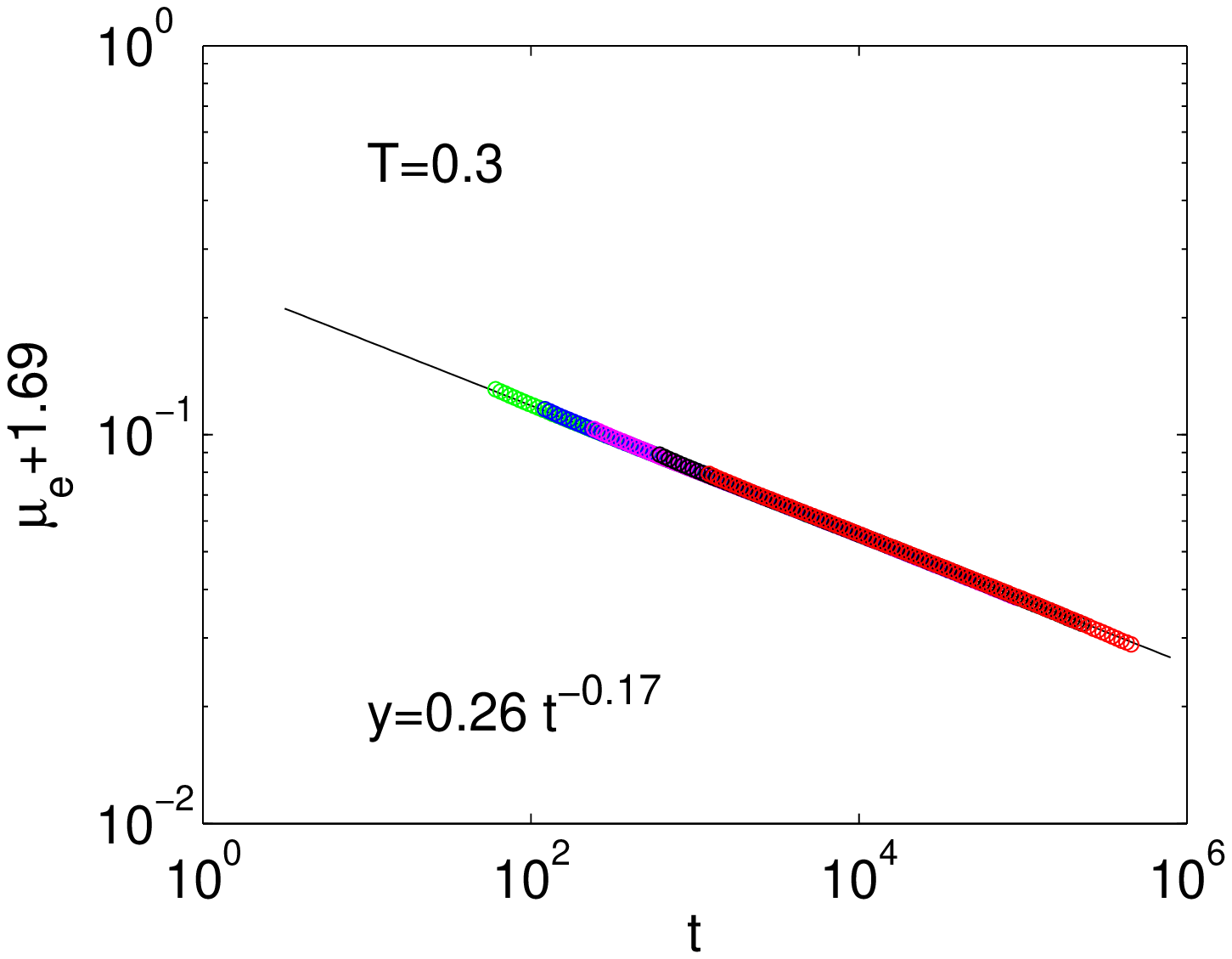} \\  
 \includegraphics[width=0.45\linewidth,height= 0.4\linewidth]{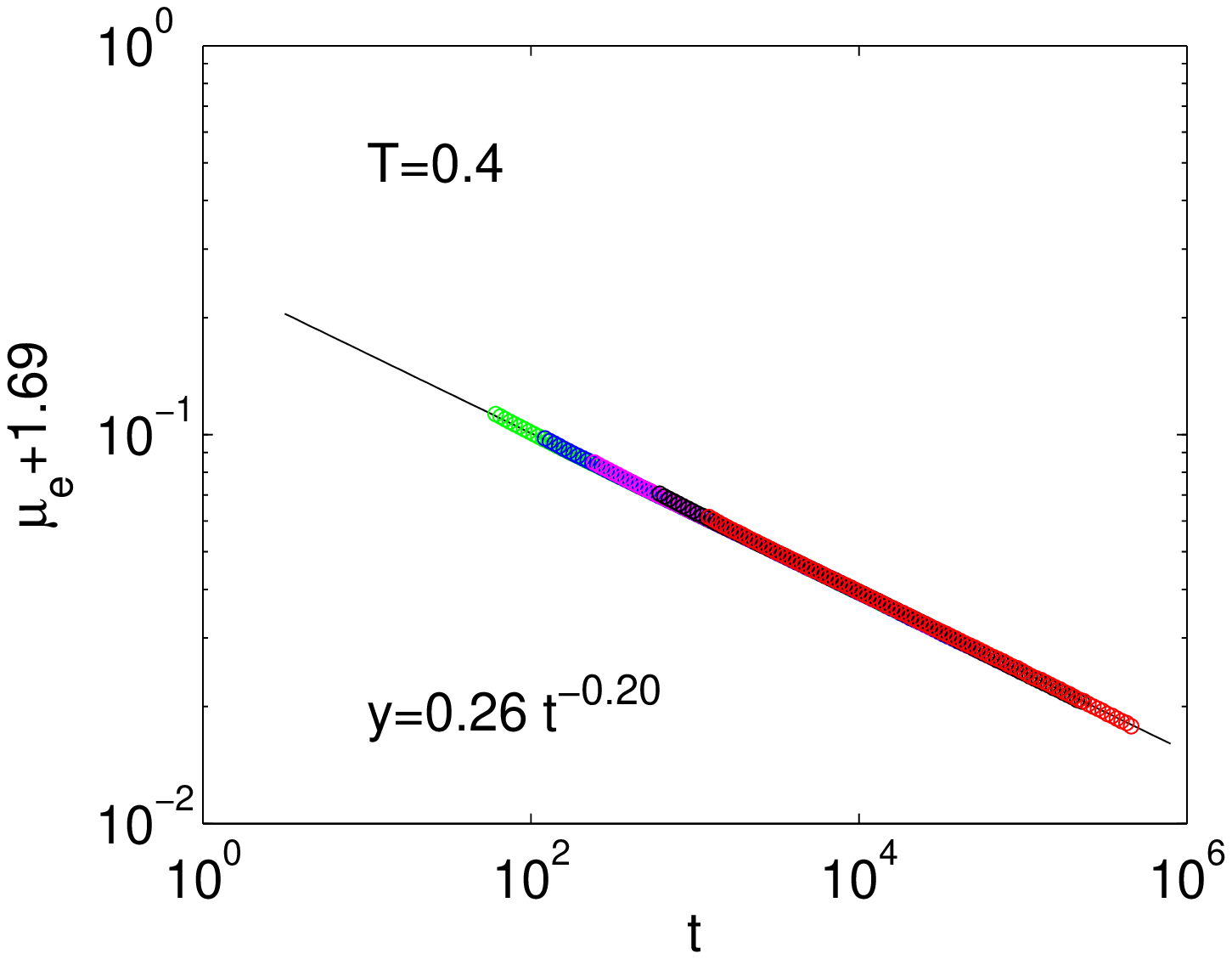}  &
 \includegraphics[width=0.45\linewidth,height= 0.4\linewidth]{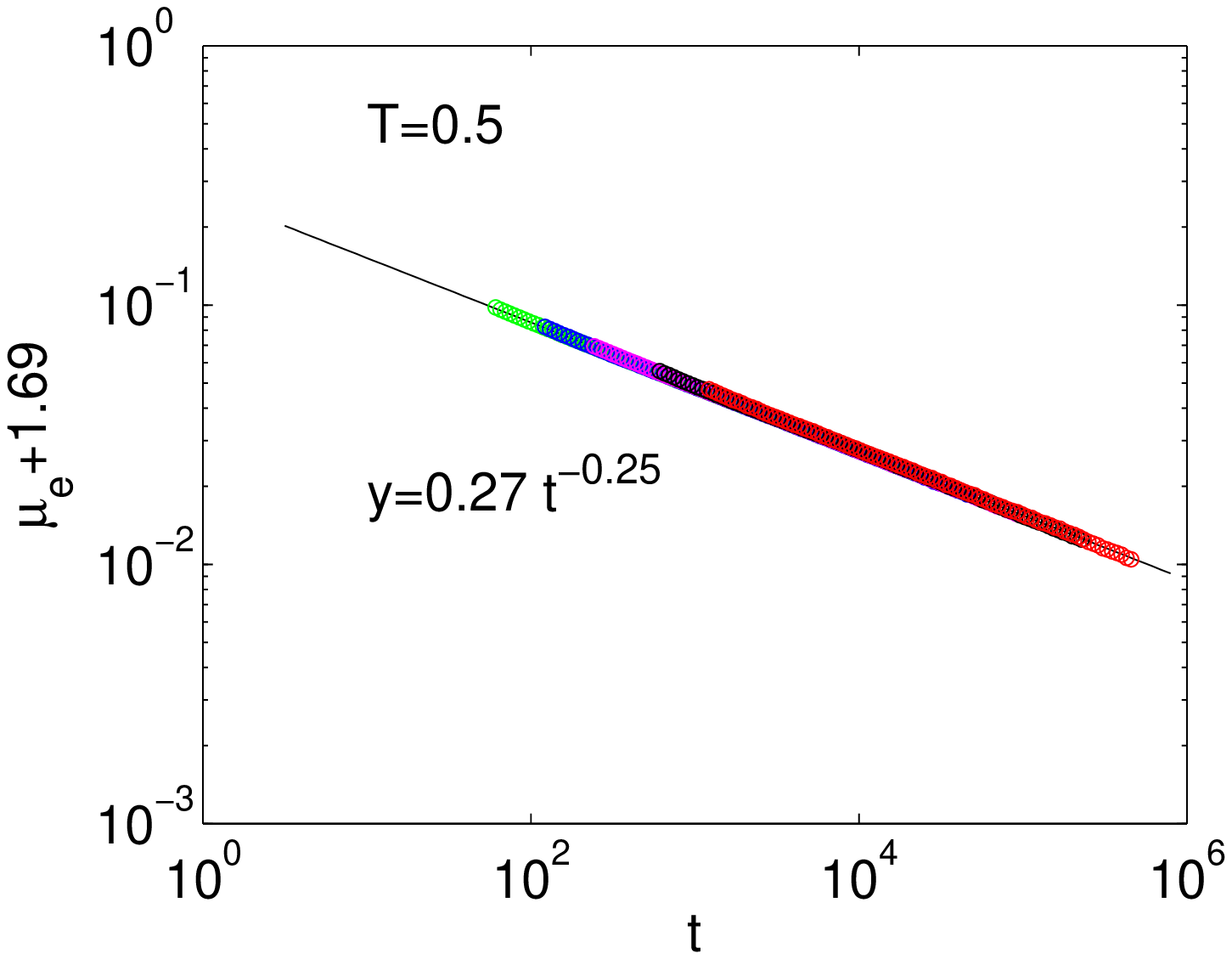}  \\  
  \includegraphics[width=0.45\linewidth,height=0.4\linewidth]{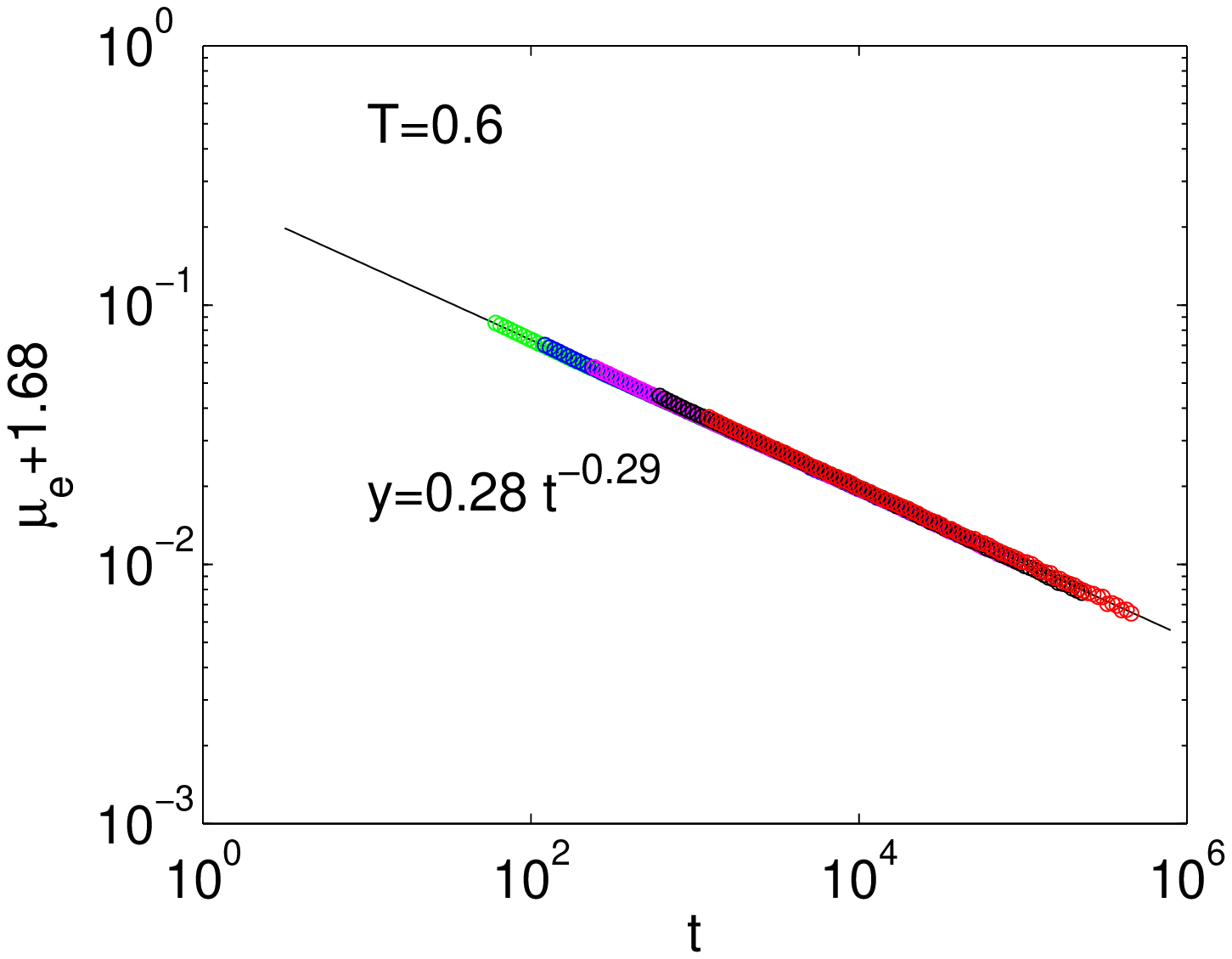}  &
  \includegraphics[width=0.45\linewidth,height= 0.4\linewidth]{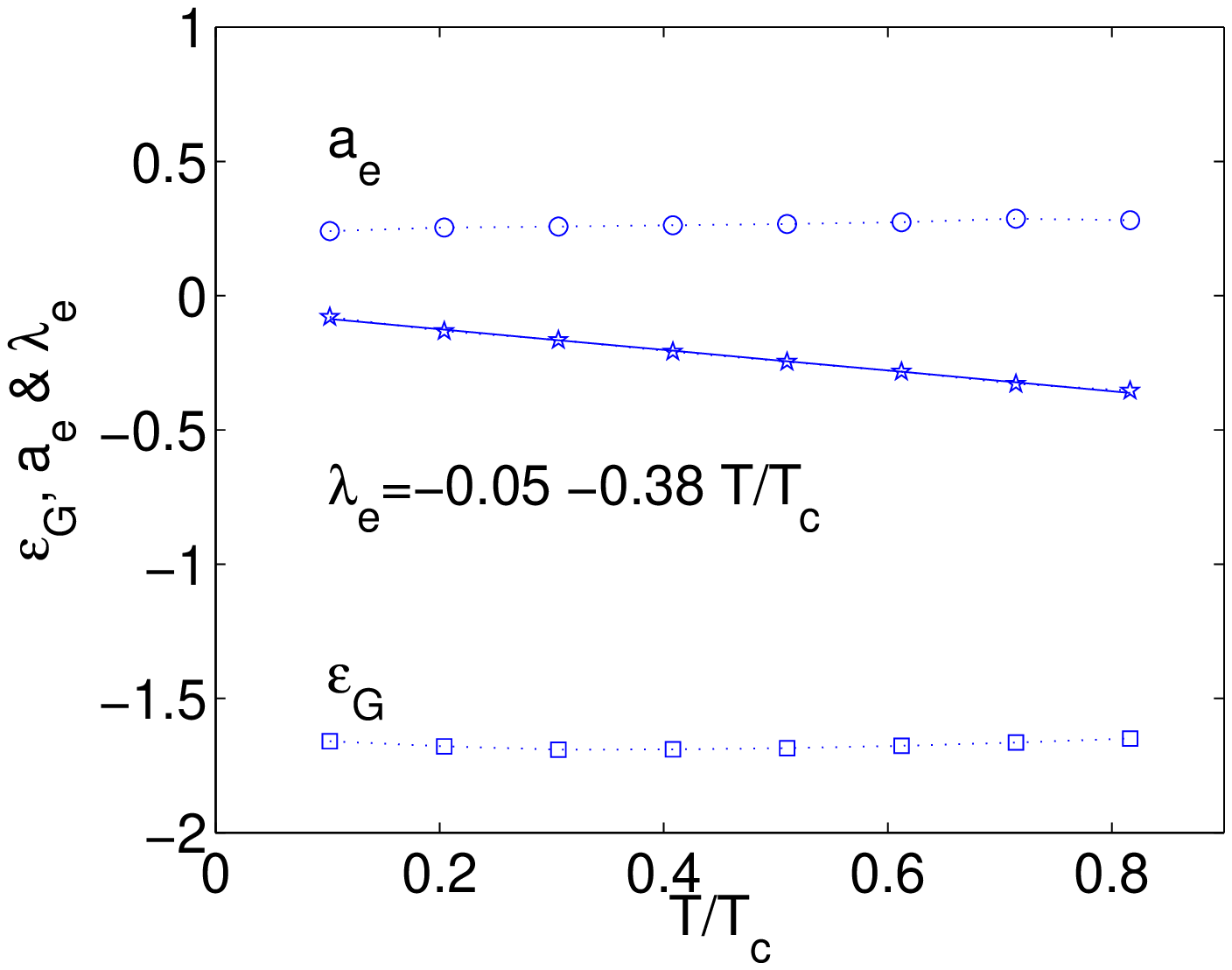} 
\end{array} 
$ 
\vspace{0.5cm} 
\caption{(Color on line) The first five panels depict, for the  temperatures indicated,  scaling plots 
of the energy per spin, with the  constant  $\epsilon_G$, subtracted.  
   The values of $t$ range from $t=t_w$ to $400 t_w$, with $t_w = 50,100,200,500$ and $1000$, 
 (green, blue, magenta, black, and red circles respectively).
 The black line is   a fit to Eq.\ref{energy_relaxation}. 
The value of  $t_w$ is  irrelevant for the energy decay. 
The last panel shows the $T$ dependence of  the parameters of Eq.~\ref{energy_relaxation} 
on  the reduced temperature $T/T_c$.  
}
\label{big_e}
\end{figure*} 
 
 \begin{figure*} 
 $
\begin{array}{cc}
\includegraphics[width=0.45\linewidth,height= 0.4\linewidth]{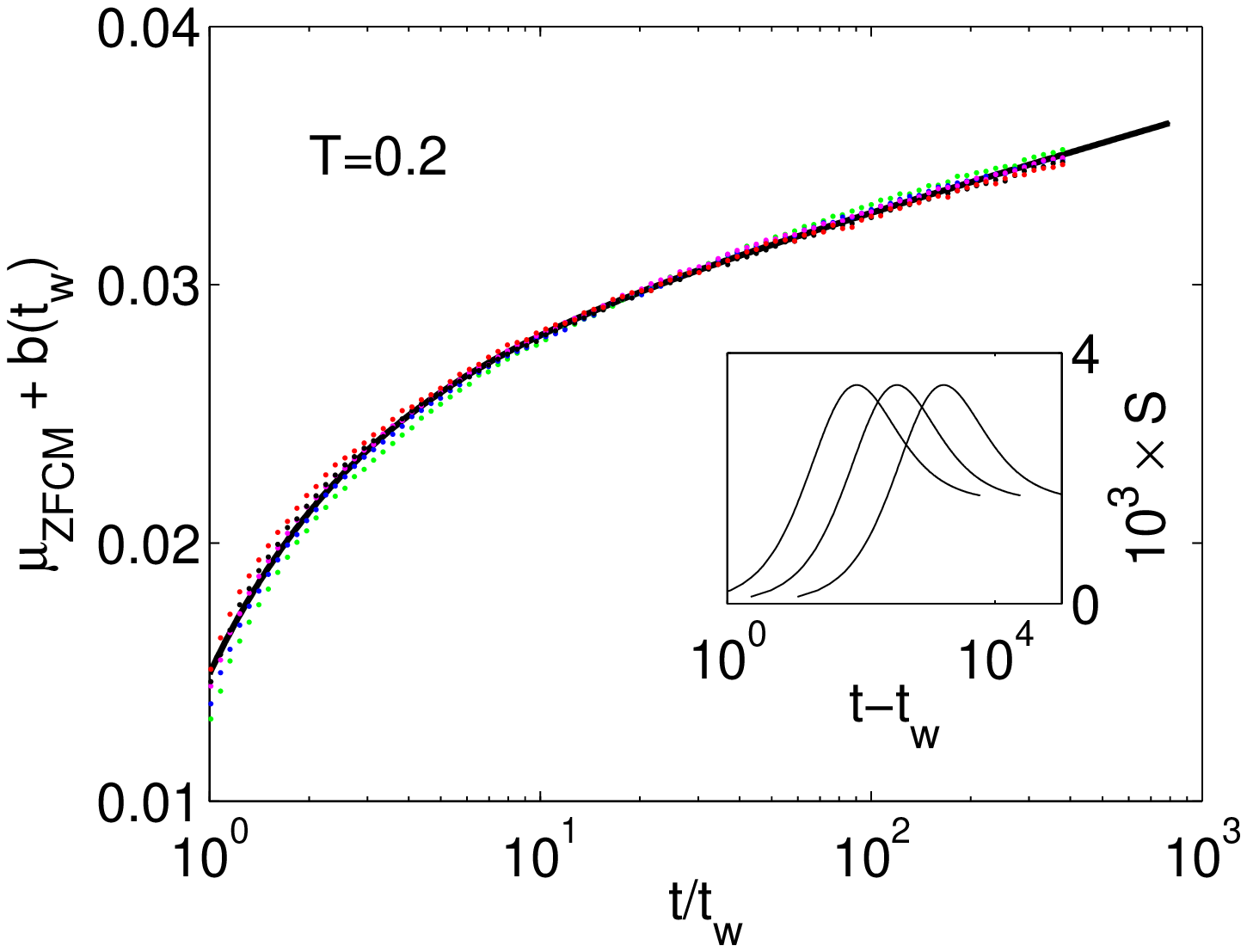}  &  
\includegraphics[width=0.45\linewidth,height= 0.4\linewidth]{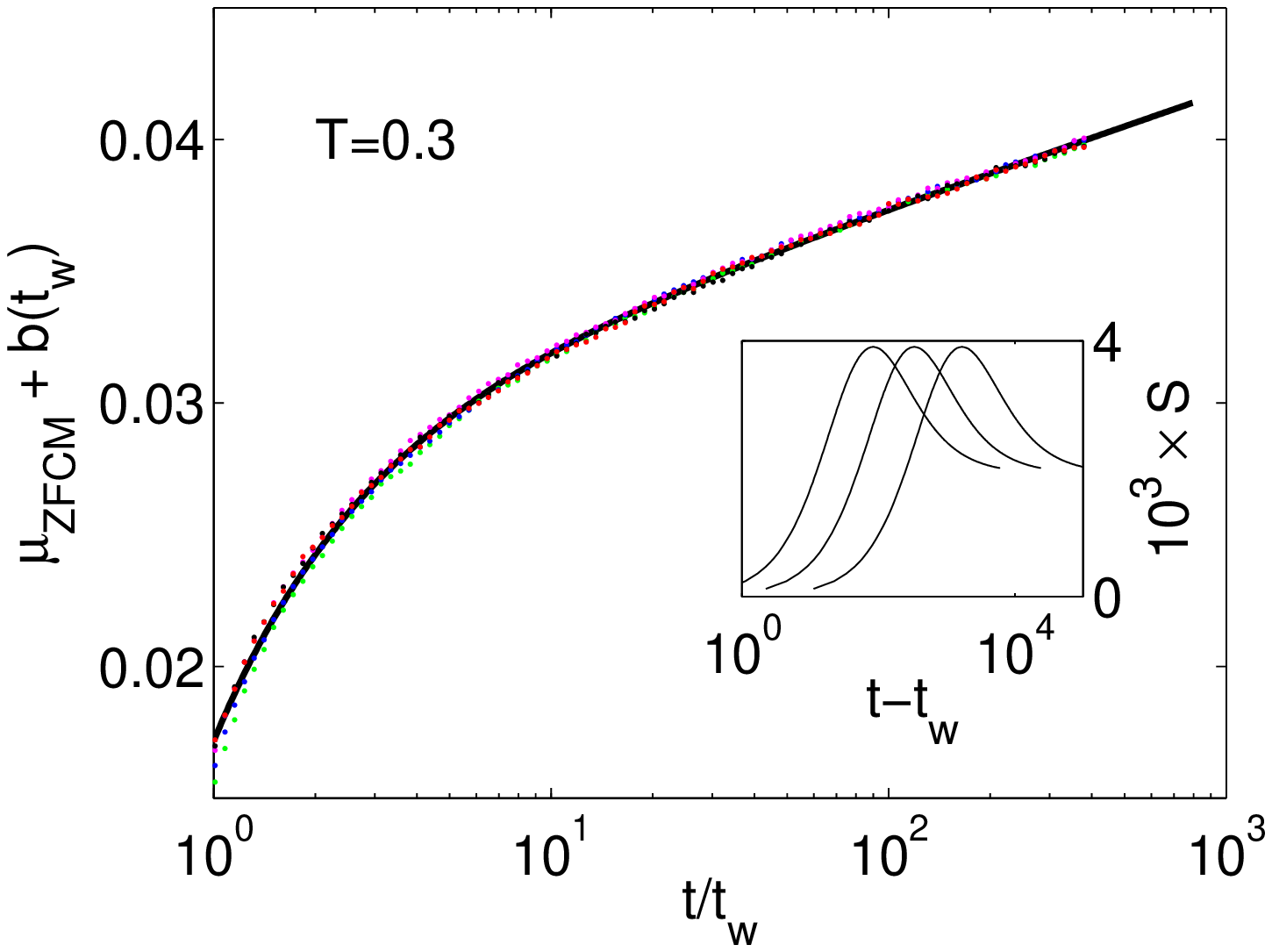} \\  
 \includegraphics[width=0.45\linewidth,height= 0.4\linewidth]{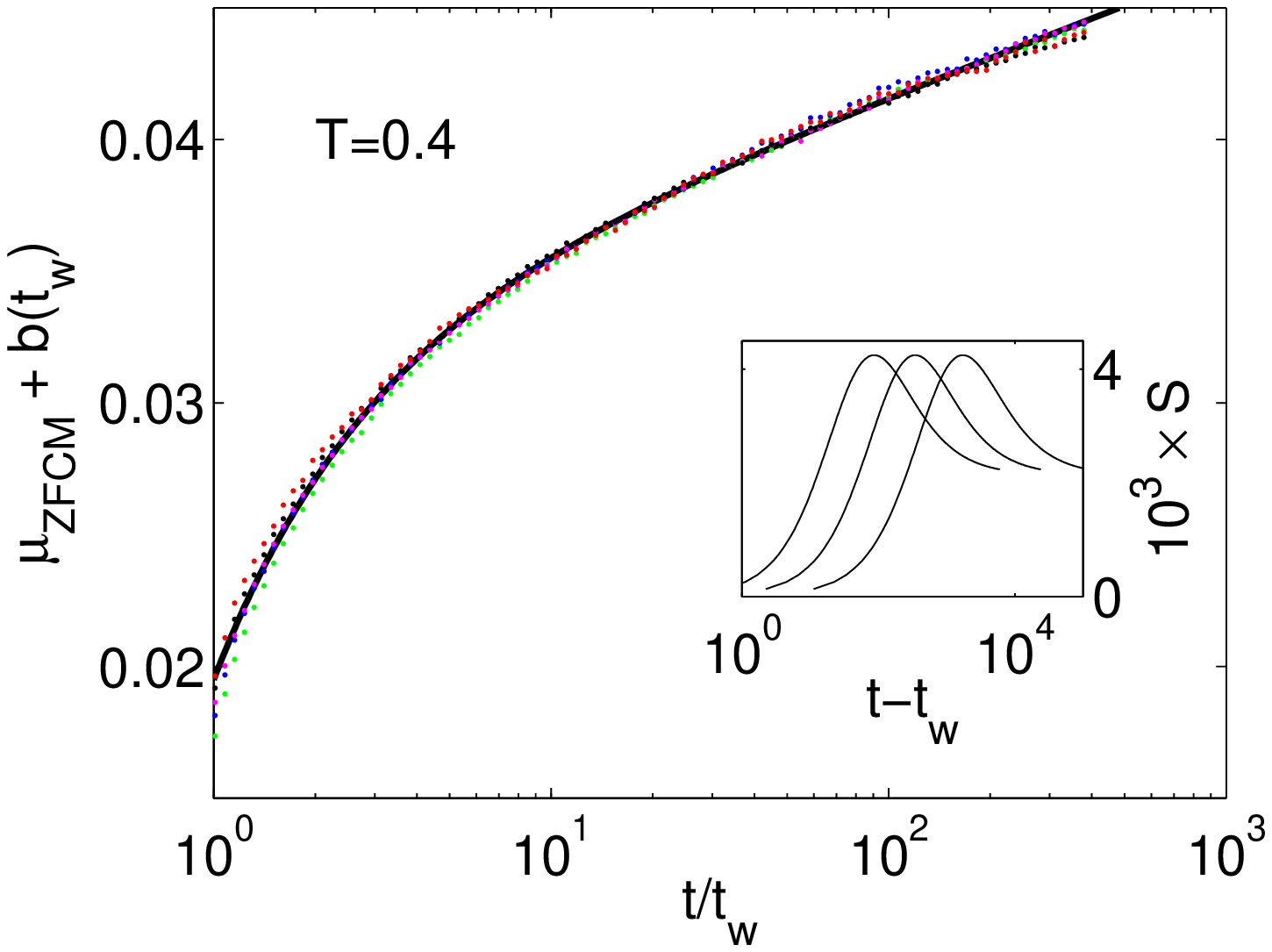}  &
 \includegraphics[width=0.45\linewidth,height= 0.4\linewidth]{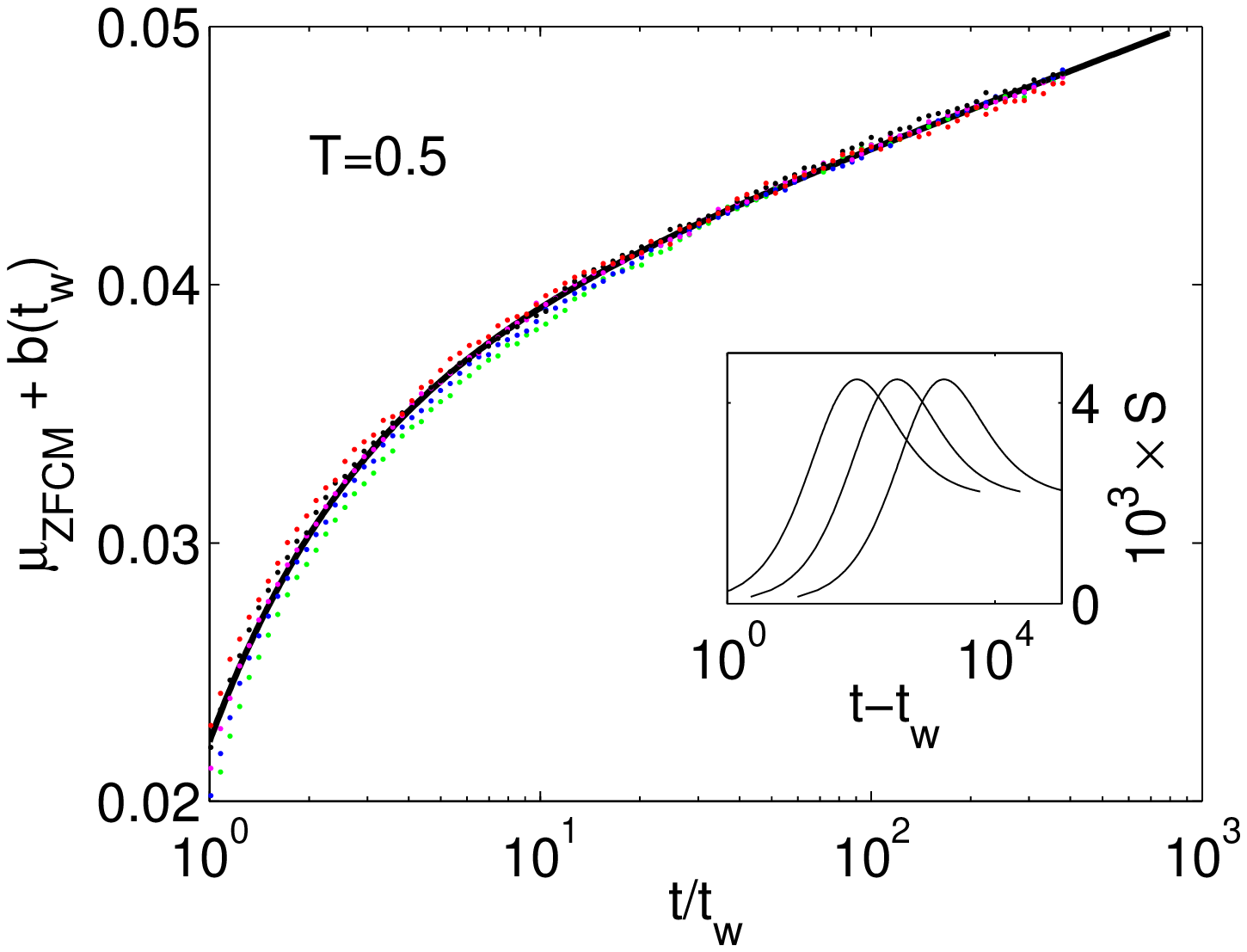}  \\  
  \includegraphics[width=0.45\linewidth,height=0.4\linewidth]{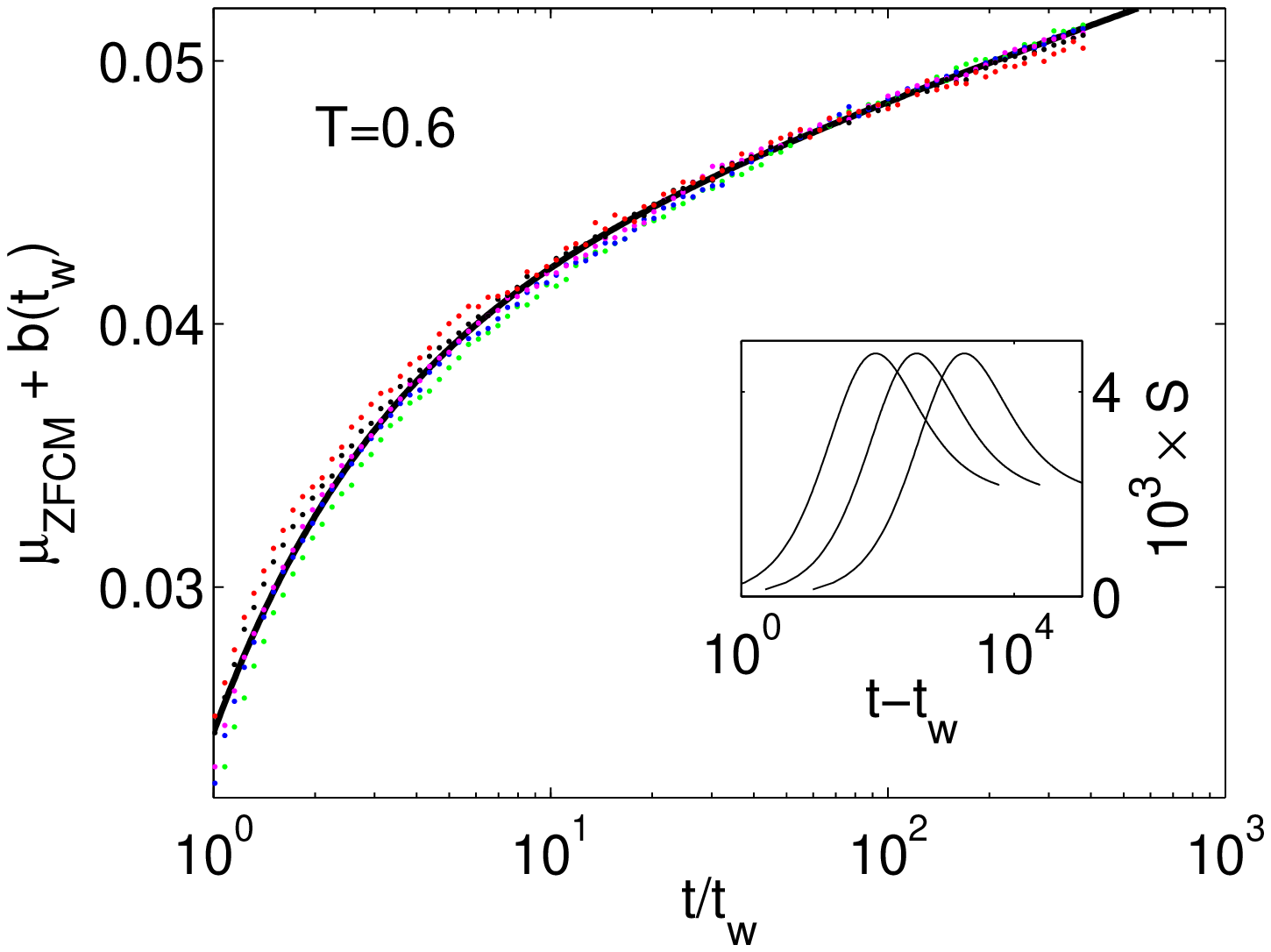}  &
  \includegraphics[width=0.45\linewidth,height= 0.4\linewidth]{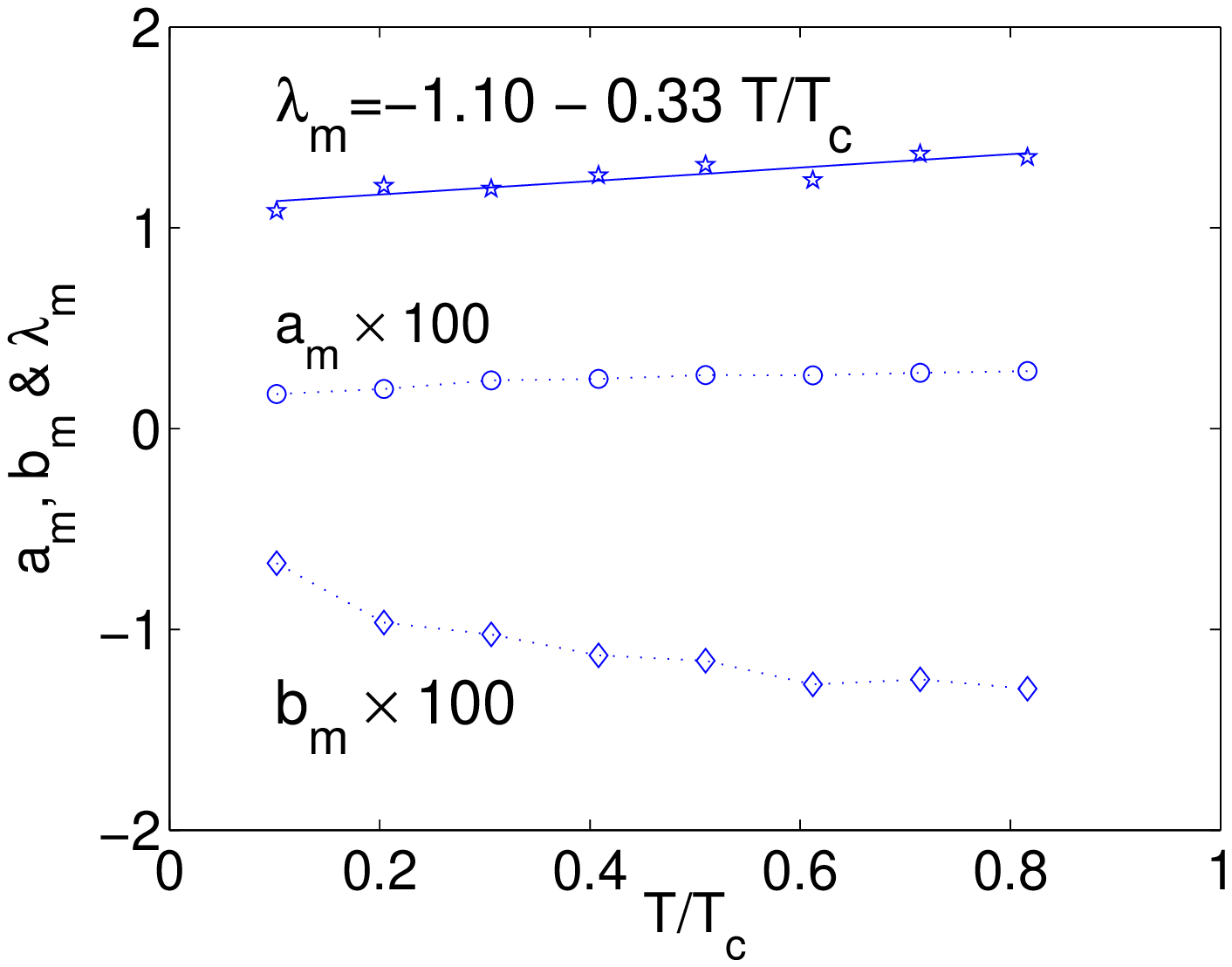} 
\end{array} 
$  
 \vspace{0.5cm} 
\caption{(Color on line) The first five panels display, for the  temperatures indicated,  scaling plots 
of the magnetization  with a small $t_w$ dependent quantity $b(t_w)$ added. 
The abscissa  is  $t/t_w$, with $t_w$ equal to  $50,100,200,500$ and $1000$.
The values of $t$ range from $t=t_w$ to $t=400  t_w$. The data are fitted to  
the sum of a logarithm and a power-law (see Eq.\ref{mag_relaxation}).
In the inserts, the relaxation rate $S$ is plotted versus the observation time $t-t_w$. 
The values of $S$ are obtained by applying Eq.~\ref{relaxation_time} 
to the fitted  functional form for $\mu_{ZFCM}$ given in Eq.\ref{mag_relaxation}.
The last panel shows the dependence of the  parameters of Eq.~\ref{mag_relaxation}   
on the reduced temperature $T/T_c$.
The negative of $\lambda_m$ is plotted for graphical reasons.
}
\label{big_m}
\end{figure*}  
The  drift of the average  energy $\mu_e$ and  average  magnetization $\mu_{ZFCM}$
 was  seen to mainly depend  on the  the number of quakes $n_I$  which fall in the relevant 
observation interval. For an interval  $(t,t')$ this number was shown to have  a Poisson 
distribution with average 
\begin{equation}
 \langle n_I \rangle = \alpha \ln(t/t').
 \label{basic}
\end{equation}
For a final check of the applicability of record dynamics
we   now compare  available  analytic   formulas~\cite{Sibani06,Sibani06a}  for the 
average energy,  magnetic response and magnetic correlation function to the spin glass data.
The subordination hypothesis used to derive the formulas fully 
neglects the  effects of 
pseudo-equilibrium fluctuations. Then, 
all  observables have a rather simple and generic 
 $n_I$ dependence---a superposition of 
exponential functions. Averaging over the distribution of $n_I$  
produces a  superposition of   power-laws terms, some of which can be further
expanded into a logarithmic and a constant term.  
 The independent   variables are $t$   and $t/t_w$  for the  energy
 and the magnetization, respectively, 
 since     all quakes  contribute to the  energy decay, while only
those falling between $t_w$ and $t$ contribute to the response.   

Because pseudo-equilibrium fluctuations are excluded,
 deviations between predicted
and observed behavior are  to be  expected. Indeed, as discussed 
later, the  linear response.
 data have  a small additive   $t_w$ dependence which 
 is beyond the reach  of the description   and which is  in most studies 
 explicitely split off as a  `stationary contribution~\cite{Picco01}.

The average energy and linear response  are calculated using 
 $4000$ independent trajectories. 
The   average  energy $\mu_e(t)$ is found to decay    toward an
(apparent) asymptotic limit  $\epsilon_G$, according to the  power-law 
\begin{equation}
\mu_e(t) - \epsilon_G = a_e t^{\lambda_e}.  
\label{energy_relaxation}
\end{equation}  
In the  first five panels of  Fig.~\ref{big_e},  the full line 
corresponds to Eq.~\ref{energy_relaxation}  and the symbols are empirical 
estimates of the average energy. Each panel shows data obtained at  the  temperature indicated.
For each $T$,  the   five data sets displayed are 
for $t_w=50, 100, 200, 500$ and $1000$ (green, blue, magenta, black and red, respectively). 
The lack of a  visible   $t_w$ dependence   shows that the magnetic contribution 
to the energy is negligible, as expected. 
The last panel of Fig.~\ref{big_e}  summarizes  the dependence 
of  $\epsilon_G$  and of the  decay exponent $\lambda_e$  on the reduced temperature $T/T_c$. 
The  estimate $T_c=0.98$~\cite{Marinari98} is  used for the critical 
temperature of the EA model. 
Interestingly,  the fitted value of  $\epsilon_G$ is nearly   independent of $T$,
and is close to the estimated ground state energy
of the EA model,~\cite{Pal96}  $e_G=-1.7003 \pm 0.008$. Eq.~\ref{energy_relaxation}
concurs with  the observation that a `putative' ground state energy
of complex optimization problems  can be guessed at early times.~\cite{Sibani90a,Tafelmayer95}
The prefactor  $a_e$ has a negligible $T$ dependence. 
In contrast,  the exponent $\lambda_e$ has a clear linear dependence    
$ \lambda_e = -0.05 -0.38 \frac{T}{T_c}$. 
 
For $\lambda_e \ln t < 1 $, the average energy  decay is 
logarithmic, $\mu_e(t) \approx \epsilon_G + a_e + a_e \lambda_e \ln (t) \ldots$,
 and the decay rate   falls off as the
inverse of the age~\cite{Sibani05,Sibani06b}.
The coefficient $a_e \lambda_e$, which  receives  its temperature dependence from $\lambda_e$,
is proportional to the average size of a quake~\cite{Sibani06b}. Even though   
quakes are strongly exothermic,  their  average size may slightly increase with  $T$.
  This  effect of thermal  activation
on the size of quakes is  analyzed (for a different model) 
 in Ref.~\cite{Sibani06b}. 

An excellent   parameterization of the  average ZFCM  time dependence  is given by
 \begin{equation} 
  \mu_{ZFCM} (t/t_w)  =  b_0  + a_m \ln \left(\frac{t}{t_w} \right)   +  b_m \left(\frac{t}{t_w} \right)^{\lambda_{m}}. 
  \label{mag_relaxation}
\end{equation}  
In the  theory,  the left-hand side  should only depend on $t/t_w$, i.e.
$b_0$ should be constant.  However, 
   a small  $t_w$ dependent term, 
 the well known stationary part of the response, is present.  To be able to 
 nevertheless plot the data versus $t/t_w$ 
 we shift them  vertically in a $t_w$ dependent fashion, i.e. we subtract the 
 stationary term.  
 The magnitude of the shift increases   with $t_w$ and with $T$,  
reaching,  at its highest,  approximately  $15$\% of  the value of $b_0$
which fits the $tw=50$ data.
 
 The   first five panels  of Fig.~\ref{big_m} show the average ZFCM (dots) and the 
 corresponding fits (lines) 
 as a function of $t/t_w$,   for  $t_w=50, 100, 200, 500$ and $1000$.
 The  simulation temperature used is indicated in each case.  
 Sets of data corresponding to different $t_w$ values are color coded  
  as done for the average  energy. The black line is given by  Eq.~\ref{mag_relaxation}, with its three 
 parameters determined  by   least square fits.   
The field switch at $t_w$ has  a `transient' effect 
  described by the  power-law decay  term:  
When    $t/t_w$ is sufficiently large,  the ZFCM increases   proportionally  
to $\ln(t/t_w)$,  and  hence at a rate decreasing   as $a_m/t$. 
In the last panel of Fig.~\ref{big_m}, 
  the decay exponent $\lambda_m$  and  the  
 constants $a_m$ and $b_m$ are plotted versus $T$. 
The full line shows the linear fit  $-\lambda_m =   1.10 + 0.33 T/T_c$,
 the dotted lines are guides to the eye.  
Importantly,    the  constant of proportionality $a_m$  is practically 
independent of  $T$, a property also shared by  $a_e$. This  
strongly indicates that  the logarithmic rate of quakes,   $\alpha$ (see Eq.~\ref{basic})
must  similarly  be $T$ independent. 
  
The     derivative of  $\mu_{ZFCM}$ with respect to 
the logarithm of the observation time, the so-called `relaxation rate',    can be written as
\begin{equation}
S(t_w,t) = (t-t_w) \frac{\partial \mu_{ZFCM}}{\partial t} =  (t-t_w)r_{ZFCM}, 
\label{relaxation_time}
\end{equation}  
where $r_{ZFCM}$ is the rate of magnetization increase. The plots of  
$S$  versus  the observation time $t_{obs}=t-t_w$ are   
produced  via   Eqs.~\ref{mag_relaxation}~and~\ref{relaxation_time} 
and dispayed in the  inserts of the first five panels  of Fig.~\ref{big_m}.
One recognizes the characteristic peak   at $t_{obs} = t_w$~\cite{Andersson92,Djurberg95}.    
The limiting value for $t\gg t_w$   is   $a_m$, precisely the asymptotic logarithmic rate of  
 increase of  the ZFC magnetization.. 

The sum of the TRM and  ZFCM 
is the field cooled  magnetization (FCM). Unlike the ZFCM, both  FCM and TRM
have a   a non-linear $H$ dependence.\cite{Djurberg95}  
The latter does not seem to affect the intermittent behavior, at least 
if  numerical simulation data and experimental data can be treated on the
same footing:  
Experimental  TRM decay data~\cite{Sibani06a} have  the same general behavior
as the ZFCM data just  discussed. However  two power-law terms, rather than   one, are   
needed to describe the, more copmlex, experimental transient. 
The   exponents have ranges similar to 
 $\lambda_m$, and a somewhat stronger  variation with $T$. The large $t/t_w$ asymptotic
 behavior is a logarithmic decay, $a \ln(t/t_w)$, where,  importantly,  the  coefficient 
  of proportionality $a$
is $T$ independent, except very close to  $T_g$.  

With an eye to the following discussion, we   recall 
 that the average configuration autocorrelation function of the EA model, normalized to
unity at $t=t_w$,  decays between times $t_w$ and $t$  as  
\begin{equation} 
  \mu_C(t/t_w)  =   \left(t/t_w\right)^{\lambda_c}; \quad  t \ge t_w,
  \label{correlation}
\end{equation}  
 where the exponent can be fitted to the linear $T$ dependence
$\lambda_c(T) = -0.25 \frac{T}{T_c}$.~\cite{Sibani06}  
The   algebraic  decay   follows from the  record dyanmics.  As   for the response, the
theoretical  arguments
neglect the effect of  quasi-equilibrium fluctuations for \mbox{$t\approx t_w$}.
Furthermore, they fail near  the final equilibration stages. 
Note that in the notation of~Ref.~\cite{Sibani06}  the ratio $t/t_w$  
is written as  $1+t/t_w$.   

\section{Discussion and Conclusions} 
In this and a  series of preceding  papers~\cite{Sibani03,Sibani04a,Sibani05,Sibani06,Sibani06b,Sibani06a},
we have argued that   non-equilibrium   events, the quakes, are  key  elements 
of intermittency and non-equilibrium aging.
The  approach     takes  irreversibility   into account at the microscopic
level,  stressing   that thermal properties alone,  e.g. free energies, are  inadequate 
to explain     non-equilibrium aging. These properties 
remain   nevertheless   important in the description,  since the record sized (positive)
energy fluctuations which 
are assumed to  trigger the quakes are drawn from an equilibrium distribution. 
As earlier discussed,~\cite{Sibani06b}  the  
energy landscape of each domain must be self-similar in order to support 
a  scale invariant statistics of record fluctuations. 
In this optics, the  scale invariance of the `local' energy landscape attached to each
domain,  rather than the scale invariance of real 
space excitations, is at  the root  the  slow relaxation behavior  of  
aging systems~\cite{Sibani89,Sibani93,Joh96}. 
 
A  different approach to 
 non-equilibrium aging  considers   a  generalization  of the
Fluctuation Dissipation Theorem (FDT)~\cite{Cugliandolo97,Herisson02,Castillo03,Calabrese05}.
 Its relation to record-dynamics is briefly considered in
below.   
Out of equilibrium, the    FDT   is never   fulfilled exactly~\cite{Diezemann05}, 
nor is it generally  possible 
to write the linear response as a function of the conjugate
autocorrelation alone.    Nevertheless, 
  for time scales $t - t_w << t_w$, the drift part of  
aging dynamics is negligible, and the FDT does, in practice, apply~\cite{Vincent96}. 
As  mentioned, conjugate response and autocorrelation functions are   naturally 
divided into   \emph{stationary} and  \emph{non-stationary}
parts, which pertain to the pseudo-equilibrium and off-equilibrium aging regimes, respectively. 
Adopting  the  standard notation $C$ for the correlation and 
$R$ for the magnetic response,  the  FDT  
reads   $(1-C(t))/T = R(t)/H$. For equilibrium data, a  plot of     $C$
versus $\chi= R/H$ yields a straight line with slope $-1/T$. For aging data,
the same  plot   produces a straight  line in the quasi-equilibrium
regime, e.g. at early times  where    $C$ is close to  one. Accordingly, a  measure  of the 
deviation from quasi-equilibrium is  the  Fluctuation Dissipation ratio (FDR)~\cite{Cugliandolo97}, 
which is defined as $X(C) = -T d \chi(C)/dC$. In the  case of the EA model,   the
relation $\ln C = \lambda_c \log(t/t_w)$, 
which follows from  inverting Eq.~\ref{correlation}, 
can be used to find $\chi(C)$ and to calculate $X(C)$. The latter is nowhere  a linear function of $C$,
which is  expected,   as the stationary fluctuation regime is excluded  from the description.  
The effective temperature~\cite{Calabrese05} 
is   usually defined   from the large $t$ and $t_w$ asymptotic limit 
of the Fluctuation Dissipation ratio. Effective temperatures may depend
on the choice of conjugate observables~\cite{Calabrese05}, and are not 
easily measured experimentally~\cite{Herisson02,Buisson03}, but offer
nevertheless a simple and appealing  characterization of aging dynamics. 

In a  record-dynamics context, 
fluctuation-dissipation like relations   arise
out of equilibrium  because   correlation and response are both subordinated to the
same quakes. Due to  the monotonicity of their   $t/t_w$ dependencies, each  of the two  can 
 can be written as a function of the other, and a  FDR can thus be constructed.   
 Asymptotically, both correlation and response   may have  an approximate  logarithmic dependence on  
$t/t_w$:  
From Eq.~\ref{mag_relaxation}, we see 
 that $R(t/t_w) \approx a_m \ln(t/t_w)$
for $t/t_w \gg 1$. 
Considering that  $\lambda_c$ vanishes for $T\rightarrow 0$, we  also obtain,  for the same range  of 
$t/t_w$ values, the inequality $|\lambda_c(T)| \ln(t/t_w) < 1$. When the inequality is fulfilled,
 Eq.~\ref{correlation} can be written as
$1-C(t/t_w) \approx  -\lambda_c(T) \ln(t/t_w) \propto R(t/t_w)$.
A relation formally similar to the FDT holds, with $T$ replaced by 
 \begin{equation}
 T_{\rm eff} = \frac{-\lambda_c H }{a_m} = \frac{T }{4 T_c (a_m/H)}.   
\end{equation}
As $a_m \propto H$, there is of course no $H$ dependence. Importantly,  $T_{\rm eff} \propto T$,
as $a_m$ is independent of  $T$. For the  parameter values  obtained from  the fits,
 $T_{eff} > T$, even though  a general argument to support the inequality is lacking at the moment.  
 
The approach  used in  this  paper and in ref.~\cite{Sibani06a}  should be generally applicable
 to  check  for the presence of record-dynamics features in 
intermittent fluctuation data. Clearly the ability to perform calorimetry experiments
is essential to directly check the temperal statistics of the quakes. In the absence of such data, one 
may assume that quakes produce large intermittent changes in other observables, and use their 
statistics instead.  
 
\section{Acknowledgments} Financial support  from the Danish Natural Sciences Research Council
is gratefully acknowledged. The bulk of the calculations was carried out on the Horseshoe Cluster
of the Danish Center for Super Computing (DCSC). 
The author is indebted to G.G. Kenning for insightful comments. 
\bibliographystyle{unsrt}
\bibliography{SD-meld}
\end{document}